# A New Lab for Measuring the Speed of Light

*Kasey Wagoner*, Princeton University, Princeton, NJ
*Richard Soden*, Princeton University, Princeton, NJ

Introduction

A typical introductory treatment of electromagnetism culminates with the investigation of Maxwell's equations, showing the beautiful connection of between the concepts covered in the many prior weeks. The lab described here is an experimental counterpart, providing a way to measure the connection between electricity, magnetism, and the speed of light. This is done using equipment that students have (likely) already explored in the lab and class.

Motivation and Background

The speed of light is such a fundamental quantity that motivating its measurement isn't necessary. In fact, it has been done in many different labs, often by measuring the time of flight of light over a given distance, for examples see [1-6]. Additionally, there are more creative ways to measure $c$, for examples see [7,8]. These are great illustrations of how first-year students can measure $c$. Here we describe a new lab to measure $c$ which as the added benefit of building on numerous topics students have been learning throughout the semester, to pull everything together into a nice, sensitive measurement of $c$.

To accomplish these goals, we wanted students to measure $c$ using devices they had already explored and were comfortable with, capacitors and solenoids (an added benefit is that these devices are readily available in introductory labs). As such, this is an *indirect* measurement of $c$ that comes from the direct measurement of the electromagnetic quantities of which it is comprised, the permittivity and permeability of free space, $\epsilon_0$ and $\mu_0$.

Background

The details of the measurement are described below, but before describing these it seems appropriate to provide context for the scenario in which we operate, though we are confident that this lab could be done in nearly any physics course. This lab was deployed near the end of the second semester of our introductory "physics for engineers" course. This allowed us to require students to perform semi-rigorous uncertainty analysis during the experiment. Additionally, we did this as a two-week lab, for a few reasons. First, our labs follow the ISLE pedagogy [9], which among other things highlight the need to measure a quantity in multiple ways to account for systematic uncertainties. Second, this gave us the time to have students vary parameters and extract measurements from best-fit values. Not all of these requirements are necessary for this measurement, but they were helpful for us to meet the many experimental learning goals, and were only possible by doing this after nearly a year of lab instruction.

Methods

This lab is done by measuring $\epsilon_0$ and $\mu_0$ independently. The results of these measurements are then combined to arrive at the measured value of $c$. Below we describe how we measured both values and provide lessons learned, which could improve the results.

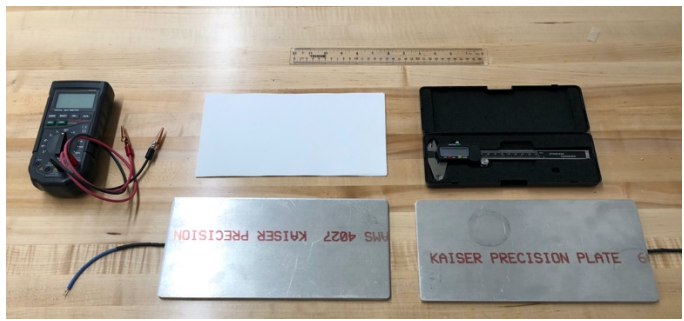
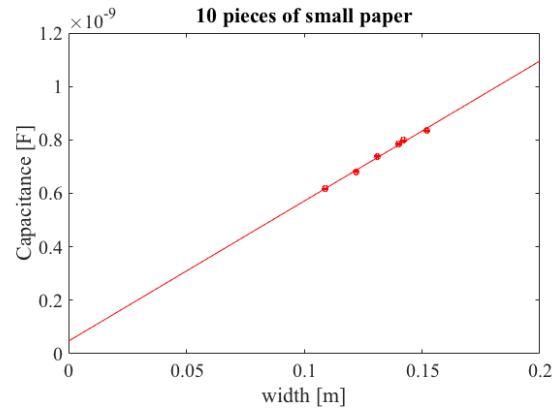

Figure 1: The equipment used to measure $\epsilon_0$ is shown. Note that the aluminum plates say "PRECISION PLATE", but there was nothing special about these plates. They were pieces of stock that were cut to the same dimensions.

Figure 2: The capacitance as a function of the capacitor width is shown for six different widths. Also shown is the best fit line. The value of $\epsilon_0$ is extracted from the slope of the best fit line. Note error bars are included but they are very small and can barely be seen.

To measure $\epsilon_0$ we used a "kitchen capacitor", a pair of large metal plates that were sandwiched around a stack of paper (Figure 1 displays our equipment). The students then use a multimeter to measure the capacitance of the arrangement. This capacitance, along with the dimensions of the capacitor and the dielectric constant of paper (this number was measured by the instructors and provided to students; for our paper it was K = $2.513 \pm 0.004$) can be used to find $\epsilon_0$. When we implemented this, students were required to measure the capacitance for different values of the capacitor geometry and then extract $\epsilon_0$ from the best fit line to the data. Figure 2 shows the data from the measurements for one student group. The permittivity value they extracted from the best fit line is $\epsilon_0 = (8.8 \pm 0.3) \times 10^{-12}$ m$^{-3}$kg$^{-1}$s$^4$A$^2$.

While this method worked well for most student groups, after deployment we realized that this measurement could be significantly improved by using the setup described in [10]. Incorporating this would allow students to make a cleaner measurement, without the need to worry about the dielectric constant of paper.

To measure $\mu_0$ we used a common solenoid which was connected to a DC power supply (Figure 3 displays our equipment). To make the measurements, students passed current through the solenoid, inserted a magnetometer into the solenoid and measured the magnetic field (for the magnetometer we used Vernier Magnetic Field Sensors: https://www.vernier.com/product/magnetic-field-sensor/, but a cheaper option would be an Adafuit magnetometer paired with an Arduino: https://www.adafruit.com/product/4022). The values of the current and magnetic field, along with the dimensions of the solenoid can be used to find $\mu_0$.

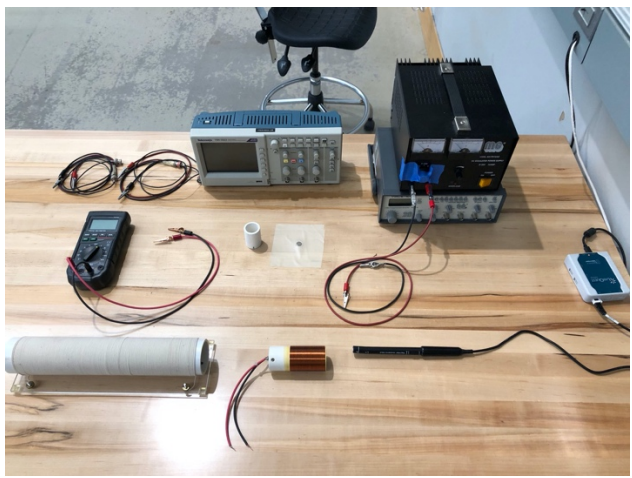

Figure 3: The equipment used to measure $\mu_0$ is shown. Note that that a signal generator and oscilloscope were provided, but they were not used in the measurements described here.

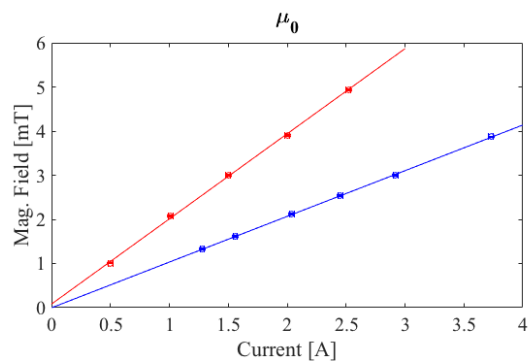

Figure 4: The magnetic field inside two solenoids as a function of the current through those solenoids is shown. For each solenoid the magnetic field is measured for multiple current values. Note error bars are included but they are very small and can barely be seen.

When we implemented this, students were required to take additional measures to improve the reliability of the experiment. The first was to measure the magnetic field for different values of current and then extract $\mu_0$ from the best fit line to the data. The second was to repeat the measurement for two different solenoids. The results of all these measurements for one student group are shown in Figure 4 (this is the same group whose $\epsilon_0$ data was shown). The two measurements resulted in values of $\mu_{0,blue} = (1.440 \pm 0.012) \times 10^{-6}$ m kg s$^{-2}$A$^{-2}$ and $\mu_{0,red} = (1.12 \pm 0.02) \times 10^{-6}$ m kg s$^{-2}$A$^{-2}$. The final result that students reported was the weighted average of these two numbers, $\mu_0 = (1.355 \pm 0.010) \times 10^{-6}$ m kg s$^{-2}$A$^{-2}$.

Putting this all together, students combine their measurements to find the speed of light
$$c = 1/\sqrt{\epsilon_0 \mu_0}.$$
The group whose data has been shown here, measured the speed of light to be $c = (2.89 \pm 0.05) \times 10^8$ m/s. While this value and quoted uncertainty isn't consistent with better measurements of $c$, it is reasonably close. The inconsistency is due to the small uncertainties quoted by these students; this provides another opportunity for exploring estimation of uncertainties.

Summary
The lab described in this manuscript has proven to be an interesting one that incorporates many instructional goals, while also highlighting the connection between electricity, magnetism, and light. Our students have enjoyed measuring this fundamental quantity with apparatus that they explored in classroom problems. Just as Maxwell's equations pull together theory which instructors have developed over the semester, this lab pulls together measurements and instruments that students have used throughout the year.